\documentclass[pra,aps,showpacs,twocolumn,twoside,superscriptaddress]{revtex4}

\usepackage{amsmath,latexsym,amssymb,revsymb}
\usepackage{theorem,graphicx}
\usepackage{bm}

\newtheorem{definition}{Definition}

\newtheorem{theorem}[definition]{Theorem}

\def\squareforqed{\hbox{\rlap{$\sqcap$}$\sqcup$}}
\def\qed{\ifmmode\squareforqed\else{\unskip\nobreak\hfil
\penalty50\hskip1em\null\nobreak\hfil\squareforqed
\parfillskip=0pt\finalhyphendemerits=0\endgraf}\fi}
\def\endenv{\ifmmode\;\else{\unskip\nobreak\hfil
\penalty50\hskip1em\null\nobreak\hfil\;
\parfillskip=0pt\finalhyphendemerits=0\endgraf}\fi}

\newcommand{\nc}{\newcommand}
\nc{\rnc}{\renewcommand} \nc{\beq}{\begin{equation}}
\nc{\eeq}{{\end{equation}}} \nc{\beqa}{\begin{eqnarray}}
\nc{\eeqa}{\end{eqnarray}} \nc{\lbar}[1]{\overline{#1}}
\nc{\bra}[1]{\langle#1|} \nc{\ket}[1]{|#1\rangle}
\nc{\ketbra}[2]{|#1\rangle\!\langle#2|}
\nc{\braket}[2]{\langle#1|#2\rangle} \nc{\proj}[1]{|
#1\rangle\!\langle #1 |} \nc{\avg}[1]{\langle#1\rangle}
\rnc{\max}{\operatorname{max}} \nc{\Rank}{\operatorname{Rank}}
\nc{\smfrac}[2]{\mbox{$\frac{#1}{#2}$}} \nc{\tr}{\operatorname{Tr}}
\nc{\ox}{\otimes} \nc{\dg}{\dagger} \nc{\dn}{\downarrow}
\nc{\cA}{{\cal A}} \nc{\cB}{{\cal B}} \nc{\cC}{{\cal C}}
\nc{\cD}{{\cal D}} \nc{\cE}{{\cal E}} \nc{\cF}{{\cal F}}
\nc{\cG}{{\cal G}} \nc{\cH}{{\cal H}} \nc{\cI}{{\cal I}}
\nc{\cJ}{{\cal J}} \nc{\cK}{{\cal K}} \nc{\cL}{{\cal L}}
\nc{\cM}{{\cal M}} \nc{\cN}{{\cal N}} \nc{\cO}{{\cal O}}
\nc{\cP}{{\cal P}} \nc{\cR}{{\cal R}} \nc{\cS}{{\cal S}}
\nc{\cT}{{\cal T}} \nc{\cX}{{\cal X}} \nc{\cZ}{{\cal Z}}
\nc{\csupp}{{\operatorname{csupp}}}
\nc{\qsupp}{{\operatorname{qsupp}}} \nc{\var}{\operatorname{var}}
\nc{\rar}{\rightarrow} \nc{\lrar}{\longrightarrow}
\nc{\polylog}{\operatorname{polylog}} \nc{\1}{{\openone}}

\nc{\RR}{{{\mathbb R}}} \nc{\CC}{{{\mathbb C}}} \nc{\FF}{{{\mathbb
F}}} \nc{\NN}{{{\mathbb N}}} \nc{\ZZ}{{{\mathbb Z}}}
\nc{\PP}{{{\mathbb P}}} \nc{\QQ}{{{\mathbb Q}}} \nc{\UU}{{{\mathbb
U}}} \nc{\EE}{{{\mathbb E}}} \nc{\id}{{\operatorname{id}}}
\nc{\oy}{\times}

\def\a{\alpha}
\def\b{\beta}
\def\g{\gamma}

\def\t{\theta}

\def\l{\lambda}

\def\r{\rho}
\def\s{\sigma}

\def\ph{\varphi}

\def\ps{\psi}

\def\Ph{\Phi}
\def\Ps{\Psi}

\begin{document}
\title{Computation of geometric measure of entanglement for pure multiqubit states}

\author{Lin Chen}
\email{cqtcl@nus.edu.sg (Corresponding~Author)} \affiliation{Centre
for Quantum Technologies, National University of Singapore, 3
Science Drive 2, Singapore 117543}

\author{Aimin Xu}
\email{xuaimin1009@yahoo.com.cn } \affiliation{Institute of
Mathematics, Zhejiang Wanli University, Ningbo 315100, China }

\author{Huangjun Zhu}
\email{zhuhuangjun@nus.edu.sg} \affiliation{Centre for Quantum
Technologies, National University of Singapore, 3 Science Drive 2,
Singapore 117543} \affiliation{NUS Graduate School for Integrative
Sciences and Engineering, Singapore 117597, Singapore}

\begin{abstract}
We provide methods for computing the geometric measure of
entanglement for two families of pure states with both experimental
and theoretical interests: symmetric multiqubit states with
non-negative amplitudes in the Dicke basis and symmetric three-qubit
states. In addition, we study the geometric measure of pure
three-qubit states systematically in virtue of a canonical form of
their two-qubit reduced states, and derive analytical formulae for a
three-parameter family of three-qubit states. Based on this result,
we further show that the W state is the maximally entangled
three-qubit state with respect to the geometric measure.
\end{abstract}

\date{\today}


\maketitle

\section{Introduction}
Quantum entanglement, which was first noted by Einstein and
Schr\"{o}dinger \cite{schrodinger35,epr35}, has been extensively
studied in the past 20 years \cite{hhh09}. In particular,
multipartite entanglement has  attracted increasing attention due to
its intriguing properties and potential applications in quantum
information processing.

The importance of multipartite entanglement can be illustrated in
two aspects. In respect of application, graph states,  prominent
examples of entangled multiqubit states, are a useful resource for
one-way quantum computation \cite{rb01} and fault-tolerant
topological quantum computation \cite{lgg09}.  Multipartite
entangled states, such as GHZ states, are essential resources for
quantum secret sharing \cite{hbb99,spc09}. In addition, multipartite
entangled states can  serve as multiparty quantum channels   in
virtue of teleportation \cite{bbc93}. In respect of theoretical
interests, multipartite states  display stronger nonlocality, one of
the key features of quantum physics \cite{ghz89, ghsz90, bbg09}.
Quantum cryptography beyond pure entanglement distillation has been
generalized to multipartite bound entangled states \cite{ah09}.
 What's more, recent
progress in experiments has made accessible more multipartite
entangled states, such as the GHZ states \cite{lzg07}, W states
\cite{pcd09}, six-photon Dicke states \cite{dicke54, wkk09,pct09}
etc. Methods for detecting such states have also been developed
\cite{kkb09}.

Given an entangled state, a natural question  to ask is how much
entanglement is contained in this state. In quantum information
theory, entanglement is usually  quantified by entanglement measures
\cite{vidal00}. An entanglement measure is an entanglement monotone,
which cannot increase under local operations and classical
communications (LOCC), and equal to zero for only classically
correlated (separable) states \cite{bds96}. Hitherto, the most
well-known entanglement measures are defined for bipartite states,
such as entanglement cost and distillable entanglement
\cite{bds96,pv07}. For pure bipartite states, there is essentially a
unique entanglement measure, the von Neumann entropy of each reduced
density matrix, which is easily computable \cite{nc00}.

For multipartite states, while a lot of entanglement measures have
been proposed \cite{hhh09,mw02,wg03,bh01}, the characterization of
multipartite entanglement is far from being complete. It is
generally difficult to calculate such measures even numerically.
Moreover, the existence of many types of inequivalent entanglement
defies a unique definition. Different entanglement measures often
induce different orders and even lead to different maximally
entangled states. For example,  the Bell state
$\ket{\Ps}=\frac{1}{\sqrt2}(\ket{00} + \ket{11})$ is the maximally
entangled state of a two-qubit system for all measures, since it
violates the Bell inequality most strongly. However, its
multipartite analog, the GHZ state
$\ket{\mathrm{GHZ}}=\frac{1}{\sqrt{2}}(\ket{000}+\ket{111})$, is
maximally entangled only under some specific entanglement measures,
such as three-tangle \cite{ckw00, twp09}. It is also maximally
entangled under any bipartition of systems \cite{ffp08}.
Nevertheless, the GHZ state consisting of more than three qubits is
not a maximally entangled state under the definition in
\cite{ffp08}, and the geometric measure of entanglement. This is one
focus of the present paper.

On the other hand, some geometrically motivated multipartite
entanglement measures have been providing us  insights on quantum
entanglement. One prominent example is the geometric measure of
entanglement (GM), \cite{bh01,wg03} which quantifies the minimum
distance between a given state and the set of product states. In
addition to providing a simple geometric picture, GM  has
significant operational meanings. It is closely related to optimal
entanglement witnesses \cite{wg03, hmm08}, and has been shown to
quantify the difficulty of multipartite state discrimination under
LOCC \cite{hmm06}. Recently, GM has also been applied to show that
most entangled states are too entangled to be useful as
computational resources \cite{gfe09}. In condensed matter physics,
GM has been utilized to study the ground state properties and to
characterize quantum phase transitions
\cite{oru08,odv08,oru08b,ow09}.

There have been extensive literatures on the quantitative
calculation of GM for both pure and mixed states
\cite{wg03,hmm08,hs09,tpt08,tkk09,pr09,chen09}. The qualitative
analysis on GM has also received much attention \cite{mmm09,hkw09}.
In addition, a few numerical methods have been developed for
computing the GM of multipartite states, such as the algorithms
presented in Refs. \cite{sb07,msb10}, which allow repeated
analytical maximization according to a subset of the parameters with
a high efficiency. However, our knowledge about GM is still quite
limited. Even for pure three-qubit states, there is no complete
analytical solution. In addition, it is still uncertain which state
is the maximally entangled with respect to GM, although the authors
of \cite{twp09} conjectured that the W state is such a candidate.
Thus it is desirable to compute GM analytically for more entangled
states, which is another focus of the present paper.

In this paper, we would like to compute the GM for several families
of multipartite pure states and determine the maximally entangled
three-qubit states with respect to GM. Throughout the paper, by
symmetric states, we mean those states which are supported on the
symmetric subspace of the whole Hilbert space.

First, we present an analytical method for computing the GM of
symmetric multiqubit states composed of Dicke states with
non-negative amplitudes by virtue of a recent simplification on GM
of symmetric states \cite{hkw09}.  Next, we analytically compute the
GM for symmetric three-qubit states. Combining with the results in
Ref. \cite{tkk09}, we provide a complete analytical solution to GM
of any symmetric pure three-qubit states. Recall that many important
multiqubit states accessible to experiments so far are symmetric,
e.g.  GHZ states \cite{lzg07}, W states \cite{pcd09}, and Dicke
states \cite{wkk09,pct09} etc. Our results may hopefully help
analyze these states in experiments.

Second, we introduce a  canonical form of pure three-qubit states
based on the canonical form of two-qubit rank-two states developed
in Ref. \cite{em02}. In virtue of  this canonical form, we study the
GM of pure three-qubit states systematically and derive explicit
analytical formulae of GM for a three-parameter family of
three-qubit states. Starting from these results, we  prove that, up
to local unitary transformations, the W state is the unique
maximally entangled pure three-qubit state with respect to GM,
confirming the conjecture made in Ref. \cite{twp09}.

The rest of the paper is organized as follows. In Sec. II, we
propose analytical methods for  computing the GM of symmetric
multiqubit states with non-negative amplitudes and that of symmetric
three-qubit states. In Sec. III, we derive analytical formulae of GM
for a three-parameter family of pure three-qubit states and prove
that the W state is the maximally entangled state under GM. We
conclude in Sec. IV.

\section{\label{sec:II}Analytical method for computing geometric measure (I): symmetric states}

The definition of GM of bipartite or multipartite pure states is
motivated by the following simple geometric idea of entanglement
quantification: the farther away from the set of separable states,
the more entangled a state is  \cite{wg03}. Given a pure state
$|\psi\rangle$ of a joint system composed of subsystems $A, B, C,
\cdots$, define $G(\ket{\psi})$ as the maximum overlap between
$|\psi\rangle$ and the set of product states, that is,
\begin{equation}
  \label{eq:gm}
  G(\ket{\ps}):=\underset{|\ph\rangle=\ket{a}\ket{b}\ket{c}\ldots}{\mathrm{max}}
  \big|\langle\ph|\ps\rangle\big|,
\end{equation}
where the normalized  one-particle states
$\ket{a},\ket{b},\ket{c},\cdots$ belong to subsystems $A, B, C,
\ldots$, respectively. $G(\ket{\psi})$ is manifestly invariant under
local unitary transformations. It obtains the maximum value 1 only
for product states and is thus an inverted entanglement measure. The
GM of a pure state is defined as follows:
\begin{equation}
  \label{eq:gmps}
  E_G(\ket{\ps}):=1-G(\ket{\ps})^2,
\end{equation}
or in another version $-2~\text{log}\ G(\ket{\ps})$ sometimes. In
this paper we will follow the definition in Eq.~(\ref{eq:gmps}). It
can be extended to the GM of mixed states by convex roof
construction \cite{wg03} according  to the same idea as in the
definition of entanglement of formation \cite{bds96}:
\begin{equation}
  \label{eq:gmms}
  E_G(\r):= \underset {\r=\sum_i p_i \proj {\ps_i}}{\mathrm{min}}
  \sum_i p_i E_G (\ket{\ps_i}).
\end{equation}
$E_G(\r) $ has been shown to be an entanglement monotone by T.-C.
Wei and P. M. Goldbart \cite{wg03}. An alternative definition of GM
for mixed states will be introduced in Sec.~\ref{sec:AP2} in a
different context.

Clearly, $E_G(\ket{\psi})$ in Eq. (\ref{eq:gmps}) is determined by
$G(\ket{\ps})$ in Eq.~(\ref{eq:gm}). From now on we focus on
$G(\ket{\ps})$ of pure states $\ket{\ps}$ and call it GM too, if
there is no confusion. For a pure bipartite state, the GM is equal
to its largest Schmidt coefficient. The problem becomes difficult
for pure  multipartite states, since there is no Schmidt
decomposition in general. The difficulty lies in the linearly
increasing number of optimization variables parametrizing  the
product states $\ket{a}\ket{b}\ket{c}\cdots$ in Eq.~(\ref{eq:gm}),
as the number of parties increases. In fact, only a few partial
results are available on this problem \cite{tpt08,wg03,tkk09,twp09}.

Recently, the authors of Ref.~\cite{hkw09} proved that, for a
symmetric pure state $|\ps^{\mathrm{sy}}\rangle$, it suffices to
consider symmetric product states in the maximization in
Eq.~(\ref{eq:gm}), that is,
\begin{equation}
  \label{eq:gmsy}
  G(\ket{\ps^{\mathrm{sy}}})=\underset {|\ph\rangle= \ket{a} \ket{a} \ket{a} \ldots}{\mathrm{max}}
  \big|\langle \ph| \ps^{\mathrm{sy}} \rangle \big|.
\end{equation}
This result can greatly simplify the calculation of GM for symmetric
states. In the rest of this section, we  derive analytical solutions
for two families of states, respectively, in virtue of this result.
In Sec.~\ref{sec:symmq}, we  analytically derive GM for symmetric
multiqubit states with non-negative amplitudes in the Dicke basis.
In Sec.~\ref{sec:sym3qA}, we derive the analytical solution of GM
for symmetric pure three-qubit states based on the previous work
\cite{tkk09}, thus solving this problem completely.

\subsection{\label{sec:symmq} symmetric multiqubit states with
non-negative amplitudes}

In this subsection, we compute the GM for pure symmetric multiqubit
states with  non-negative amplitudes in the Dicke basis. More
explicitly, we investigate the $N$-qubit state
\begin{equation}
  \label{eq:symq}
  \ket{\ps^{\mathrm{symq}}}:=\sum^N_{m=0}a_m\ket{m,N},
\end{equation}
where $a_m\geq0$, and $\ket{m,N}$ is the Dicke state \cite{dicke54}
defined as
\begin{equation}
  \label{eq:dicke}
  \ket{m,N}:={N \choose
   m}^{-1/2} \sum_k P_k \ket{\overbrace{1,...,1}^{m},\overbrace{ 0,...,0}^{N-m}},
\end{equation}
where $P_k$s denote the set of all permutations of the spins. By
definition, Dicke states are symmetric; so the state
$\ket{\ps^{\mathrm{symq}}}$ is also symmetric, and we can apply
Eq.~(\ref{eq:gmsy}) to computing its GM. Let
$\ket{a}=\cos\a\ket{0}+e^{i\t}\sin\a \ket{1}$ with
$\a\in[0,\frac{\pi}{2}]$ and $\t\in[0,2\pi]$; then the GM of the
state in Eq.~(\ref{eq:symq}) reads
\begin{align}
  \label{al:gmsymqnonnegative}
  &G(\ket{\ps^{\mathrm{symq}}}) \nonumber\\
  &=\underset{|\ph\rangle=\ket{a}\ket{a}\ket{a}\ldots}{\mathrm{max}}
  \bigg|\sum^N_{m=0}{N \choose
   m}^{1/2}a_m\cos^{N-m}\a\sin^{m}\a
  e^{-im\t}\bigg|,\nonumber\\
  &\leq\underset{|\ph\rangle=\ket{a}\ket{a}\ket{a}\ldots}{\mathrm{max}}
  \sum^N_{m=0}\bigg|{N \choose
   m}^{1/2}a_m\cos^{N-m}\a\sin^{m}\a
  \bigg|,\nonumber\\
  &=\underset{\a\in[0,\frac{\pi}{2}]}{\mathrm{max}}
  \sum^N_{m=0}{N \choose
   m}^{1/2}a_m\cos^{N-m}\a\sin^{m}\a,
\end{align}
where the equality holds when $\t=0$.
Equation~(\ref{al:gmsymqnonnegative})  contains only one variable
$\a$, so one can easily find out the maximum. For example, let
$x=\tan\a$, then one can convert $G^2(\ket{\ps^{\mathrm{symq}}})$
into a rational fraction ${A(x)}/{B(x)}$, where $A(x)$ and $B(x)$
are both polynomials on $x$. By calculating its derivative we can
find out the maximum in Eq.~(\ref{al:gmsymqnonnegative}) explicitly.

A similar idea can be applied to calculating the GM of any symmetric
multi-qudit state with nonnegative amplitudes in the generalized
Dicke basis;  again the number of free variables can be reduced by
half. Recently, the additivity of GM of  states with non-negative
amplitudes was proved, i.e., $G(\ket{\a} \ox \ket{\b}) =
G(\ket{\a})~G(\ket{\b})$ ~\cite{zch10}. So we can compute the GM of
$\ket{\ps^{\mathrm{symq}}} \ox \ket{\b}$ if we know the GM of
$\ket{\b}$ too.

Unfortunately, the present method does not apply to arbitrary
symmetric multiqubit states, e.g., those states in
Eq.~(\ref{eq:symq}) having negative or complex amplitudes. For more
complicated states, numerical methods are indispensable for
computing their entanglement measures, see, for example,
Refs.~\cite{sb07,msb10}.

\subsection{\label{sec:sym3qA} symmetric three-qubit states}

In this subsection, we  compute the GM of symmetric pure three-qubit
states. Such states can always be converted into the following form
with suitable  local unitary operations \cite{tkk09}:
\begin{equation}
  \label{eq:sytqgeneral}
  \ket{\Ph} = g\ket{000} + t(\ket{011} + \ket{101} + \ket{110})+ e^{i\g} h
  \ket{111},
\end{equation}
where $g, t, h \geq 0$ and $\g \in [-\frac{\pi}{2}, \frac{\pi}{2}]$.
So it suffices to calculate the GM for the state $\ket{\Ph}$.

Analytical formula of the GM is already known if  at least one of
the three parameters $g, t, h$ is vanishing, or
$\g=0,\pm\frac{\pi}{2}$ \cite{wg03,tpt08,hmm08,tkk09,notation1}.
Hence, we can focus on the family of states with
\begin{align}
  \label{al:sytq}
  g, t, h > 0, \;\; g^2+3t^2+h^2=1, \;\; \g \in (-\frac{\pi}{2},0) \cup
  (0,\frac{\pi}{2}).\nonumber\\
\end{align}

The authors of Ref.~\cite{tpt08}  have reduced the task of computing
the GM of the state $|\Ph\rangle$ to solving the following system of
equations of the three variables $\ph, \t, \l$ (see also appendix
A):
\begin{align}
  \label{al:sytqaxiom1}
  2 h t \cos\g + 2 t (g+t) \sin\t \cos\ph -& \nonumber \\
  2 h t \cos\g \cos\t = & \l \sin\t \cos\ph,\\
  \label{al:sytqaxiom2}
  2 h t \sin\g - 2 t (g-t) \sin\t \sin\ph -&  \nonumber \\
  2 h t \sin\g \cos\t =& \l \sin\t \sin\ph, \\
  \label{al:sytqaxiom3}
  (g^2-t^2) (1+\cos\t) -h^2 (1-\cos\t) -&  \nonumber\\
  2 h t \cos\g \sin\t \cos\ph - 2 h t \sin\g \sin\t \sin\ph  =& \l
  \cos\t.
\end{align}
For each root $(\ph_j, \t_j)$ of
Eqs.~(\ref{al:sytqaxiom1})--(\ref{al:sytqaxiom3}), we can obtain a
GM candidate of the state $\ket{\Ph}$ via the following formula,
according to Eq.~(\ref{al:gsquare}) in appendix A,
\begin{align}
  \label{al:gmsytq}
  G_j^2(\ket{\Ph}) &=
 \frac18 \big[3 - 2t^2 + 4(1 - 2h^2 - 4t^2) \cos\t_j
  \nonumber\\
   &+ (1 - 6 t^2) \cos 2\t_j + 4 g t \cos 2\ph_j \sin^2 \t_j  \nonumber \\
   &+ 32 h t \cos (\g-\ph_j) \cos \frac{\t_j}{2}
  \sin^3 \frac{\t_j}{2}\big];
\end{align}
the GM is the maximum over all the GM candidates:
\begin{eqnarray}
G^2(\ket{\Ph})=\underset{ j} {\mathrm{max} }\,G_j^2(\ket{\Ph}).
\end{eqnarray}

We  shall solve Eqs.~(\ref{al:sytqaxiom1})--(\ref{al:sytqaxiom3}) in
two cases separately; the second case consists of three subcases. In
each case we  obtain one or a few GM candidates by computing
Eq.~(\ref{al:gmsytq}) with the solutions to the system of equations.

\textbf{Case 1}. Suppose $\t = 0$; then the phase $\ph$ does not
play any role, and Eqs.~(\ref{al:sytqaxiom1}) and
(\ref{al:sytqaxiom2}) become identities, while
Eq.~(\ref{al:sytqaxiom3}) determines $\l$. In this case we get a GM
candidate via Eq.~(\ref{al:gmsytq}) as follows:
\begin{equation}
  \label{eq:gmsytq1}
  G_1^2(\ket{\Ph}) = g^2.
\end{equation}

\textbf{Case 2}  To satisfy
Eqs.~(\ref{al:sytqaxiom1})--(\ref{al:sytqaxiom3}), the roots
$\ph=k\frac{\pi}{2}$ for $k=0,1,2,3$ lead to $\t=0$, which is
already discussed. Moreover, $\t=\pi$ cannot be a legal solution of
Eqs.~(\ref{al:sytqaxiom1}) or~(\ref{al:sytqaxiom2}), so this choice
is excluded. Hence, it remains to solve
Eqs.~(\ref{al:sytqaxiom1})--(\ref{al:sytqaxiom3}) under the
assumption that
\begin{align}
  \label{al:assump}
  \ph &\in (0, \frac{\pi}{2}) \cup (\frac{\pi}{2}, \pi) \cup
  (\pi, 3\frac{\pi}{2}) \cup (3\frac{\pi}{2}, 2\pi),
  \quad \t \in (0, \pi).
\end{align}

From Eqs.~(\ref{al:sytqaxiom1}--\ref{al:sytqaxiom3}), we can
determine $\l$ as a function of $\t$ and $\ph$. Inserting this
solution into Eq.~(\ref{al:sytqaxiom1}) and
Eq.~(\ref{al:sytqaxiom2}), we can obtain two equations about $\t$
and $\ph$: $\text{eq}1(\ph,\t)=0$ and $\text{eq}2(\ph,\t)=0$,
respectively. This further implies that either
\begin{equation}
  \label{eq:tan1}
 \tan \frac{\t}{2}=\frac{g} {h\csc 2\ph \sin (\g - \ph)},
\end{equation}
or
\begin{equation}
  \label{eq:tan2}
 \tan \frac{\t}{2}=\frac{-t} {h\csc 2\ph \sin (\g + \ph)}.
\end{equation}
Combining either of them and $\text{eq}1(\ph,\t)=0$ can lead to a
set of solutions.

\textbf{Case 2.1} There is a simple solution $\tan \ph=\frac{t+g}
{t-g} \tan {\g}$. The variable $\t$ can be determined via either
Eq.~(\ref{eq:tan1}) or~(\ref{eq:tan2}), which lead to an identical
result in this case. Inserting this solution into
Eq.~(\ref{al:gmsytq}), we get another GM candidate:
\begin{align}
  \label{al:gmsytq21}
  G_2^2(\ket{\Ph}) = g^2 - \frac {(g^2 - t^2)^3} {t^2-2t^4+g^2-6g^2 t^2-2g t h^2 \cos
  2\g}.
\end{align}

\textbf{Case 2.2}  By combining Eq.~(\ref{eq:tan1}) and
$\text{eq}1(\ph,\t)=0$, we can get two polynomial equations:
\begin{align}
  \label{al:sytq22}
  \sum^4_{i=0} c_{1i}(g,t,h,\g) \cos^i 2\ph=0,\nonumber\\
  \sum^4_{i=0} c_{2i}(g,t,h,\g) \cos^i 2\ph=0,
\end{align}
as well as Case 2.1, which has already been handled. Since
Eqs.~(\ref{al:sytq22}) are quartic equations on $\cos 2\ph$, we can
analytically derive their roots. We may obtain up to 16 different
phases $\ph \in [0,2\pi]$. The variable $\t$ can then be determined
via Eq.~(\ref{eq:tan1}). Hence, we can derive up to 16 GM candidates
$G_j(\ket{\Ph})$ for $j=3,\ldots,18$ via Eq.~(\ref{al:gmsytq}). This
 differs a bit from Case 1 and Case 2.1, where only one GM
candidate is given respectively.

\textbf{Case 2.3} Similar to Case 2.2,  by combining
Eq.~(\ref{eq:tan2}) and $\text{eq}1(\ph,\t)=0$, we can get two
quartic polynomial equations on $\cos 2\ph$, which are analytically
solvable. Again we may get up to 16 GM candidates $G_j(\ket{\Ph})$
for $j=19,\ldots,34$. This finishes the discussion of Case 2.

Now we have all the GM candidates $G_j(\ket{\Ph})$ for
$j=1,\ldots,34$. The maximum of them is exactly the GM of the state
$\ket{\Ph}$ in Eq.~(\ref{eq:sytqgeneral}).

For the convenience of the readers, here we repeat the main steps
for deriving the GM of the symmetric three-qubit state $\ket{\Ph}$.

{\bf Step 1.} Compute $G_1(\ket{\Ph})$ and $G_2(\ket{\Ph})$ via
Eqs.~(\ref{eq:gmsytq1}) and ~(\ref{al:gmsytq21}), respectively; two
GM candidates can be obtained.

{\bf Step 2.} Compute $G_j(\ket{\Ph})$ for $j=3,\ldots,18$ via
Eq.~(\ref{al:gmsytq}) with roots ($\ph_j, \t_j$)  of
Eqs.~(\ref{eq:tan1}) and ~(\ref{al:sytq22}); up to 16 GM candidates
can be obtained.

{\bf Step 3.} Similar to  Step 2, with the roots ($\ph_j, \t_j $)
for $j=19,\ldots,34$ of Eq.~(\ref{eq:tan2}) and quartic equations
similar to Eq.~(\ref{al:sytq22}), up to 16 GM candidates can be
obtained via Eq.~(\ref{al:gmsytq}).

{\bf Step 4.} The maximum of all 34 GM candidates is exactly the GM
of the state $\ket{\Ph}$.

In conclusion, we have provided a method for analytically deriving
the GM of the symmetric three-qubit states in
Eq.~(\ref{eq:sytqgeneral}) with $\g \in (-\frac{\pi}{2}, 0) \cup (0,
\frac{\pi}{2})$. The special cases $\g = 0, \pm \frac{\pi}{2}$ have
been addressed in Ref. \cite{tkk09}. Calculation shows that our
result approaches their result when $\gamma$ approaches these
special values. Hence, we can now compute the GM of any symmetric
pure three-qubit states.

\section{\label{sec:AP2} Analytical method for computing geometric measure (II):
maximal entangled states among pure three-qubit states}

In this section, we introduce a  canonical form of pure three-qubit
states based on the canonical form of two-qubit rank-two states
developed in Ref. \cite{em02}. By virtue of this canonical form, the
GM of  pure three-qubit states is studied systematically. In
particular, we derive analytical formulae of GM for the family of
pure three-qubit states one of whose rank-two two-qubit reduced
states is the convex combination of the maximally entangled state
and its orthogonal pure state within the rank-two subspace. Based on
these results, we prove that the W state is the maximally entangled
three-qubit state with respect to GM, confirming the  conjecture in
Ref. \cite{twp09}.

Our approach builds on Theorem 1 in Ref. \cite{jung08}, which states
that the GM of an $n$-partite pure state $|\psi\rangle$ is
determined by any of its ($n-1$)-partite reduced states $\rho$, that
is,
$$G^2(|\psi\rangle)=g(\rho).$$
Here $g(\rho)$ is an alternative definition of geometric measure and
has nothing to do with the parameter $g$ introduced in
Eq.~(\ref{eq:sytqgeneral}):
\begin{eqnarray}
g(\rho)=\underset{\rho_1,\ldots,\rho_{n-1}}{\max}\;
\mathrm{tr}\bigl[\rho(\rho_1\otimes\cdots\otimes\rho_{n-1})\bigr],\label{gmmix}
\end{eqnarray}
where $\rho_1,\ldots,\rho_{n-1}$ are pure single-particle states,
namely $\r_i = \proj{a_i}$.  In addition, to any closest product
state $\rho_1\otimes\cdots\otimes\rho_{n-1}$ of $\rho$, there
corresponds a unique closest product state of $|\psi\rangle$ with
$\rho_1\otimes\cdots\otimes\rho_{n-1}$ as a reduced state. A closest
product state of $\rho$ is any pure product state
$\rho_1\otimes\cdots\otimes\rho_{n-1}$ that maximizes
Eq.~(\ref{gmmix}). From the definition, $g(\rho)$ is a convex
function of $\rho$; this property will be frequently resorted to
later.

Note that for a mixed state $\rho$, $g(\rho)$ is not the standard
definition of the GM of $\rho$ (see the first paragraph of
Sec.~\ref{sec:II}). Nevertheless, this alternative definition is
useful for computing the GM of any purification of $\rho$
\cite{jung08}. It has also many applications of its own, such as
constructing optimal entanglement witnesses \cite{wg03, hmm08} and
quantifying the difficulty of state discrimination under LOCC
\cite{hmm06, hmm08}.

\subsection{\label{sec:canonical}  Canonical form of two-qubit
rank-two states}

In this section we introduce a  canonical form of pure three-qubit
states based on the canonical form of two-qubit rank-two states
developed in Ref. \cite{em02} and set the notations useful in later
discussions.

For a pure three-qubit state, each two-qubit reduced state lies on a
rank-two subspace of the two-qubit Hilbert space. Up to local
unitary transformations, the projector $\Sigma_0$ onto a general
rank-two subspace can be specified by just two parameters $\gamma_1,
\gamma_2$ \cite{em02}:
\begin{eqnarray}
\label{eq:r2projector}
\Sigma_0&=&\frac{1}{2}(1+u\sigma_3+v\tau_3+z_1\sigma_1\tau_1+z_2\sigma_2\tau_2),\nonumber\\
u&=&\cos\gamma_1\cos\gamma_2,~~v=\sin\gamma_1\sin\gamma_2,\nonumber\\
z_1&=&\sin\gamma_1\cos\gamma_2,~~z_2=\cos\gamma_1\sin\gamma_2,\nonumber\\
&&\mbox{with}\quad \frac{1}{2}\pi\geq\gamma_1\geq\gamma_2\geq 0,
\end{eqnarray}
where $\sigma_1,\sigma_2,\sigma_3$ are the Pauli operators for the
first qubit and $\tau_1,\tau_2,\tau_3$ are that for the second
qubit. Interchange of the two qubits leads to
$\gamma_1\rightarrow\frac{\pi}{2}-\gamma_2$,
$\gamma_2\rightarrow\frac{\pi}{2}-\gamma_1$. So without loss of
generality, we can assume $\gamma_1+\gamma_2\leq\frac{\pi}{2}$,
$\gamma_2\leq\frac{\pi}{4}$.

Any rank-two state supported on $\Sigma_0$ can be written as
follows:
\begin{eqnarray}
\rho_{\mathrm{rk}2}&=&\frac{1}{2}(\Sigma_0+x_1\Sigma_1+x_2\Sigma_2+x_3\Sigma_3),
\label{eq:ranktwoCF}
\end{eqnarray}
where $\Sigma_1, \Sigma_2,\Sigma_3$ are the Pauli operators for the
rank-two subspace \cite{em02}:
\begin{eqnarray}\label{eq:Pauli}
\Sigma_1&=&\frac{1}{2}(\sin\gamma_1\sigma_1+\cos\gamma_2\tau_1+\sin\gamma_2\sigma_1\tau_3+\cos\gamma_1\sigma_3\tau_1),\nonumber\\
\Sigma_2&=&\frac{1}{2}(\sin\gamma_2\sigma_2+\cos\gamma_1\tau_2+\sin\gamma_1\sigma_2\tau_3+\cos\gamma_2\sigma_3\tau_2),\nonumber\\
\Sigma_3&=&\frac{1}{2}(v\sigma_3+u\tau_3-z_2\sigma_1\tau_1-z_1\sigma_2\tau_2+\sigma_3\tau_3),
\end{eqnarray}
and $(x_1,x_2,x_3)$ (satisfying $x_1^2+x_2^2+x_3^2\leq 1$) is  the
Bloch vector of $\rho_{\mathrm{rk}2}$.

Local unitary symmetry and complex conjugation symmetry play an
important role in determining the behavior of
$g(\rho_{\mathrm{rk}2})$  and in simplifying its calculation.
According to Eqs.~(\ref{eq:r2projector}) and (\ref{eq:Pauli}),
simultaneous local unitary transformation $\sigma_3\otimes \tau_3$
flips the sign of $\Sigma_1, \Sigma_2$, while leaving $\Sigma_0,
\Sigma_3$ invariant, that is,
\begin{eqnarray}
\sigma_3\otimes \tau_3\Sigma_{0,3}\sigma_3\otimes
\tau_3&=&\Sigma_{0,3},\nonumber\\
\sigma_3\otimes \tau_3\Sigma_{1,2}\sigma_3\otimes
\tau_3&=&-\Sigma_{1,2};
\end{eqnarray}
under this transformation, the Bloch vector of $\rho_{\mathrm{rk}2}$
changes as follows: $(x_1,x_2,x_3)\rightarrow(-x_1,-x_2,x_3)$.
Complex conjugation flips the sign of $\Sigma_2$, that is
\begin{eqnarray}
\Sigma_j^*=(-1)^{\delta_{j,2}}\Sigma_j,
\end{eqnarray}
where $\delta_{j,2}$ is the Kronecker $\delta$ function; under this
transformation, the Bloch vector of  $\rho_{\mathrm{rk}2}$  changes
as follows: $(x_1,x_2,x_3)\rightarrow(x_1,-x_2,x_3)$ . As a
consequence of these symmetries, any reasonable entanglement measure
is equal for the four states $\rho_{\mathrm{rk}2}$ with Bloch
vectors $(\pm x_1,\pm x_2,x_3)$, respectively. Without loss of
generality, we can assume $x_1, x_2\geq 0$.

If $\gamma_1=\gamma_2$,  the simultaneous local unitary
transformation $\mathrm{e}^{-\mathrm{i}
\theta\sigma_3\otimes\tau_3}$ rotates the Bloch vector of
$\rho_{\mathrm{rk}2}$ around the $x_3$ axis, so
$g(\rho_{\mathrm{rk}2})$ is rotationally invariant about the $x_3$
axis. If $\gamma_2=0$, the local unitary transformation
$\mathrm{e}^{-\mathrm{i} \theta\tau_1}$ rotates the Bloch vector
around the $x_1$ axis, so $g(\rho_{\mathrm{rk}2})$ is rotationally
invariant about the $x_1$ axis.

Up to local unitary transformations, there is a one-to-one
correspondence between pure three-qubit states and rank-two
two-qubit states. Hence, the canonical form of rank-two two-qubit
states provides a canonical form of pure three-qubit states.
Moreover, due to Theorem 1 in Ref. \cite{jung08} and the arguments
given above, computing the GM of pure three-qubit states can be
reduced to computing $g(\rho_{\mathrm{rk}2})$ of the family of
canonical rank-two two-qubit  states in Eq.~(\ref{eq:ranktwoCF})
with $0\leq \gamma_2\leq\gamma_1\leq\frac{\pi}{2}$,
$\gamma_2+\gamma_1\leq \frac{\pi}{2}$, $0\leq x_1,x_2\leq1$ and
$-1\leq x_3\leq1$. With this background, we can now study the GM of
pure three-qubit states systematically.

\subsection{\label{sec:generic} $g(\rho_{\mathrm{rk}2})$ of general two-qubit rank-two states}
In this subsection we reduce the task of computing
$g(\rho_{\mathrm{rk}2})$ for general two-qubit rank-two  states to a
maximization problem which involves only two free variables. The
number of free variables is further reduced to one for states
$\rho_{\mathrm{rk}2}$ with $x_2=0$.

Let
 $\rho_1=\frac{1}{2}(1+\bm{s}_1\cdot\bm{\sigma})$ and $\rho_2=\frac{1}{2}(1+\bm{s}_2\cdot\bm{\tau})$ be two pure qubit
states with Bloch vectors $\bm{s}_1=(a,b,c)$ and
$\bm{s}_2=(a_2,b_2,c_2)$, respectively. Straightforward calculation
shows that
\begin{eqnarray}
&&\mathrm{tr}(\rho_{\mathrm{rk}2}\rho_1\otimes\rho_2)=\frac{1}{4}(1+a
x_1\sin\gamma_1+bx_2\sin\gamma_2\nonumber\\
&&{}+c\cos\gamma_1\cos\gamma_2+cx_3\sin\gamma_1\sin\gamma_2+\bm{w}\cdot\bm{s}_2),
 \label{trace0}
\end{eqnarray}
where
\begin{eqnarray}
\bm{w} &&=\left(
             \begin{array}{c}
               a(\cos\gamma_2\sin\gamma_1- x_3\cos\gamma_1\sin\gamma_2) \\
              b(- x_3\cos\gamma_2\sin\gamma_1+\cos\gamma_1\sin\gamma_2) \\
               b x_2\sin\gamma_1+a x_1\sin\gamma_2\ \\
             \end{array}
           \right)^T \nonumber\\
           &&{}+\left(
             \begin{array}{c}
               c x_1\cos\gamma_1+x_1\cos\gamma_2 \\
               x_2\cos\gamma_1+cx_2\cos\gamma_2 \\
               c x_3+x_3\cos\gamma_1\cos\gamma_2+\sin\gamma_1\sin\gamma_2\ \\
             \end{array}
           \right)^T.  \label{trace1}
\end{eqnarray}
Given $\rho_{\mathrm{rk}2}$ and $\rho_1$,  the trace in
Eq.~(\ref{trace0}) is maximized when $\bm{s}_2$ is parallel to
$\bm{w}$. According to Eq.~(\ref{gmmix}), we have
\begin{eqnarray}
g(\rho_{\mathrm{rk}2})&&=\underset{\rho_1 \rho_2}{\max}\;
\mathrm{tr}(\rho_{\mathrm{rk}2}\rho_1 \ox \rho_2)
\nonumber\\
&&=\frac{1}{4}\underset{a^2+b^2+c^2=1}{\max}f(a,b,c), \nonumber\\
f(a,b,c)&&=(1+a x_1\sin\gamma_1+bx_2\sin\gamma_2
+c\cos\gamma_1\cos\gamma_2\nonumber\\
&&\quad{}+cx_3\sin\gamma_1\sin\gamma_2+|\bm{w}|),\label{fv1}
\end{eqnarray}
where $|\bm{w}|$ denotes the  Euclidian norm of $\bm{w}$. Thus we
have reduced the task of computing $g(\rho_{\mathrm{rk}2})$ to that
of maximizing the function $f(a,b,c)$ on the unit sphere determined
by $a^2+b^2+c^2=1$, which involves only two free variables. The
contours of $f$ are in general some quadratic surfaces. At the
maximum of $f$ over the unit sphere, the contour is generally
tangent to the sphere. This geometric picture is useful in
visualizing the closest product states.

Further simplification is possible for states $\rho_{\mathrm{rk}2}$
with $x_2=0$. Assuming $x_2=0$ and $x_1\geq0$ (recall that we  only
need  to consider the case $0\leq x_1,x_2\leq1$ due to consideration
on symmetry, see Sec.~\ref{sec:canonical}); then $f(a,b,c)$ is an
even function of $b$ according to Eq.~(\ref{fv1}). In addition,
$f(|a|,b,c)\geq f(-a,b,c)$ and  $f(a,\sqrt{1-a^2},c)$ is
nondecreasing with $a$ for $a\geq0$. So the maximum of $f(a,b,c)$
can be obtained in the parameter subspace satisfying $a\geq0$,
$b=0$. Moreover,  the maximum can only be found in this subspace if
$x_1>0$. Thus the calculation of $g(\rho_{\mathrm{rk}2})$ can be
reduced to the optimization problem over the single variable $c$.

For rank-two subspace with $\gamma_1=\gamma_2$ or $\gamma_2=0$, this
simplification is applicable to all states $\rho_{\mathrm{rk}2}$,
since it is enough to calculate $g(\rho_{\mathrm{rk}2})$ for states
with $x_2=0$ due to the symmetry discussed in
Sec.~\ref{sec:canonical}.

\subsection{\label{sec:W}The W state is the maximally entangled state with respect to the geometric measure}

According to the discussion in Sec.~\ref{sec:canonical}, to
determine the maximally entangled states of three-qubit  with
respect to GM, it is enough to determine the global minimum of
$g(\rho_{\mathrm{rk}2})$ over the set of canonical two-qubit
rank-two states. Due to the convexity (cf. Eq.~(\ref{gmmix})) and
the symmetry of $g(\rho_{\mathrm{rk}2})$, given $\gamma_1,\gamma_2,
x_3$, the minimum of  $g(\rho_{\mathrm{rk}2})$ (as a function of
$x_1$ and $x_2$) is obtained at $x_1=x_2=0$. So the global minimum
of $g(\rho_{\mathrm{rk}2})$ can be obtained at states with this
property. Recall that the state $\rho_{\mathrm{rk}2}$ with
$x_1=x_2=0$, $x_3=-1$ is the maximally entangled state in the
rank-two subspace \cite{em02}. Hence, these states are convex
combination of the maximally entangled state and its orthogonal pure
state within the rank-two subspace. They are interesting for a
couple of reasons. First, the two-qubit reduced states of many
important pure three-qubit states, such as the W state, GHZ state,
are among this family of states. Second, from the result on this
family of states and that on pure states, both an upper bound and a
lower bound for $g(\rho_{\mathrm{rk}2})$ of any state in each
rank-two subspace can be obtained by virtue of the convexity and
symmetry properties of $g(\rho_{\mathrm{rk}2})$.

Hence, in order to find the maximally entangled three-qubit state
with respect to GM, it suffices to investigate
$g(\rho_{\mathrm{rk}2})$ of the states $\rho_{\mathrm{rk}2}$ with
$x_1=x_2=0$. After some elementary algebra (see Appendix B), we
derive a simple analytical formula of $g(\rho_{\mathrm{rk}2})$ of
this family of states. To emphasize its explicit dependence on the
three parameters $x_3, \gamma_1,\gamma_2$,  we write
$g(x_3,\gamma_1,\gamma_2)$ for $g(\rho_{\mathrm{rk}2})$ .
\begin{eqnarray}
\label{gm1}
&&g(x_3,\gamma_1,\gamma_2)=\nonumber\\
&&\left\{
\begin{array}{cc}
  \frac{(1-x_3)[1+\cos(\gamma_1+\gamma_2)]}{4},&\mathrm{I}\vspace{0.5ex}\\
  \frac{(1-x_3^2)\sin\gamma_1\cos\gamma_2(\cos\gamma_2\sin\gamma_1-x_3\cos\gamma_1\sin\gamma_2)}
{-2\bigl[x_3^2-(\cos\gamma_2\sin\gamma_1-x_3\cos\gamma_1\sin\gamma_2)^2\bigr]},& \mathrm{II}\vspace{0.5ex} \\
  \frac{(1+x_3)[1+\cos(\gamma_1-\gamma_2)]}{4}, &\mathrm{III}
\end{array}
\right.
\end{eqnarray}
where I, II, III denote three intervals, ${\mathrm{I}: -1\leq
x_3\leq x_3^{(3)}}$, $\mathrm{II}:x_3^{(3)}< x_3< x_3^{(4)}$,
$\mathrm{III}: x_3^{(4)}\leq
  x_3\leq1 $. Here
$x_3^{(3)}$  and $x_3^{(4)}$ are given by{\small
\begin{eqnarray}
&&x_3^{(3,4)}(\gamma_1,\gamma_2) :=\nonumber\\
&&\frac{-\sin\gamma_1\{\pm\sin\gamma_1+[\cos\gamma_1\cos\gamma_2+(\sin\gamma_1)^2]\sin\gamma_2\}}
{1+\cos\gamma_1\{\cos\gamma_2-\sin\gamma_2[\cos\gamma_1\sin\gamma_2+(\sin\gamma_1)^2\tan\gamma_2]\}},\nonumber\\
 \label{interval2}
\end{eqnarray}}
and obey the inequalities: $-1\leq x_3^{(3)}\leq0\leq
x_3^{(4)}\leq1$.
\begin{figure}
  \includegraphics[width=6.5cm]{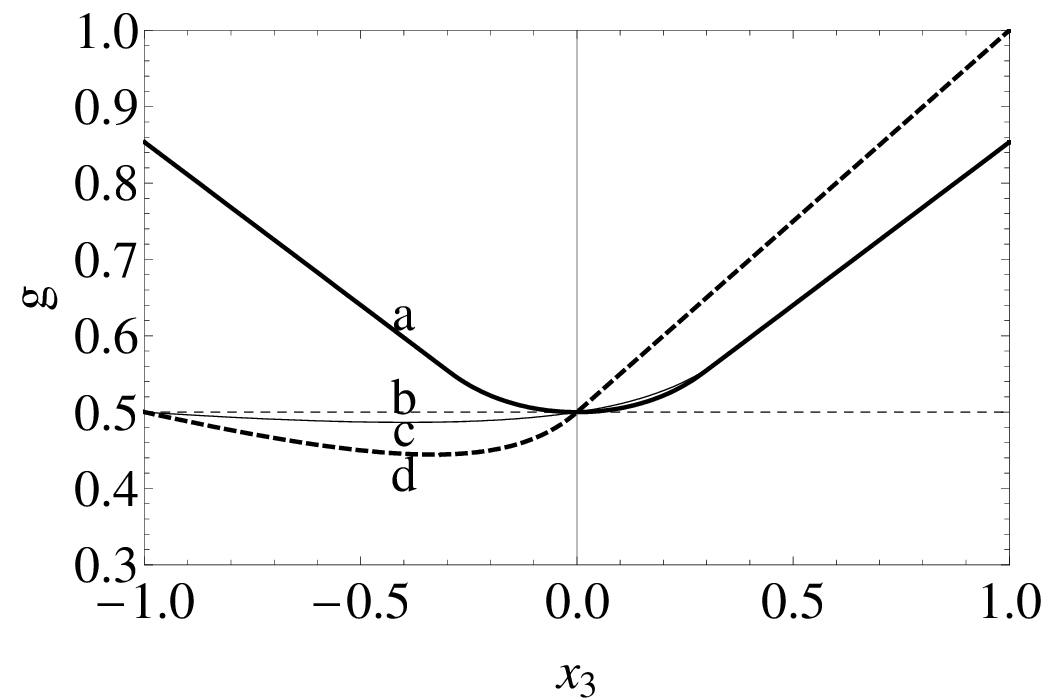}\\
  \caption{\label{fig:gm0}  $g(x_3,\gamma_1$, $\gamma_2)$ as a function of $x_3$ for several different two-qubit rank-two
  subspaces.
   (a) $\gamma_1=\frac{\pi}{4}$, $\gamma_2=0$, (b) $\gamma_1=\frac{\pi}{2}$, $\gamma_2=0$,
  (c) $\gamma_1=\frac{3\pi}{8}$, $\gamma_2=\frac{\pi}{8}$, and (d) $\gamma_1=\gamma_2=\frac{\pi}{4}$.
  Curve (a) and curve (c) coincide in a large interval, because
  $\gamma_1-\gamma_2=\frac{\pi}{4}$ for both the rank-two subspaces, see Eq.~(\ref{gm1}).}
\end{figure}
\begin{figure}
  \includegraphics[width=6cm]{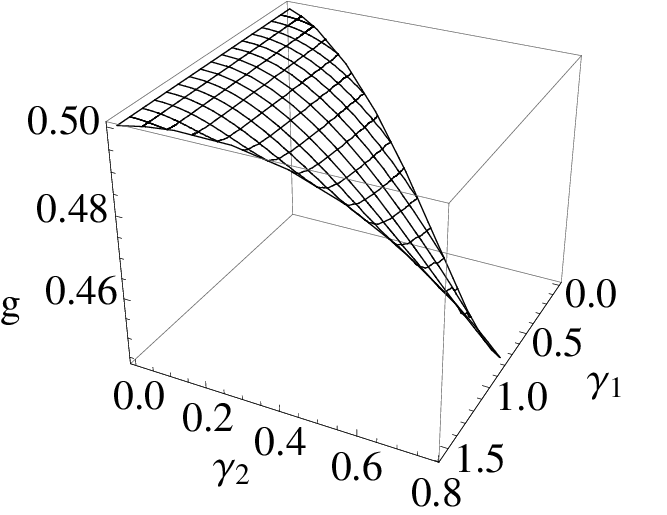}\\
  \caption{\label{fig:gm3}  The minimum of $g(x_3,\gamma_1,\gamma_2)$ over $x_3$ in each rank-two subspace;
  the global minimum is obtained at $\gamma_1=\gamma_2=\frac{\pi}{4}$.}
\end{figure}

Figure~\ref{fig:gm0} shows $g(x_3,\gamma_1,\gamma_2)$ as a function
of $x_3$ for several different values of $\gamma_1,\gamma_2$.
$g(x_3,\gamma_1,\gamma_2)$ is equal to
$\frac{1}{2}\bigl[1+\cos(\gamma_1\mp\gamma_2)\bigr]$ at $x_3=\pm1$.
This is consistent with the well-known result on GM of two-qubit
pure states, which is a function of the concurrence; recall that the
concurrence is equal to $\sin(\gamma_1\mp\gamma_2)$ for the  two
states with $x_3=\pm1$ \cite{em02}.  $g(x_3,\gamma_1,\gamma_2)$ is
equal to $\frac{1}{2}$ at $x_3=0$, which is independent of the other
two parameters. This observation implies that
$G^2(|\psi\rangle)=\frac{1}{2}$ for any pure three-qubit state which
is maximally entangled under some bipartite partition; a perfect
example is the GHZ state. Once the value of
$g(x_3,\gamma_1,\gamma_2)$ is specified at the three points
$x_3=\pm1, 0$, the value of $g(x_3,\gamma_1,\gamma_2)$ at a generic
point can roughly be estimated by interpolation, keeping the
convexity of $g(x_3,\gamma_1,\gamma_2)$ with respect to $x_3$ in
mind. This simple picture is very useful in understanding the
dependence of $g(x_3,\gamma_1,\gamma_2)$ on various parameters, and
why the W state is the maximally entangled state with respect to GM,
as we shall see shortly.

It is interesting to note that, when $x_3>x_3^{(4)}$ or
$x_3<x_3^{(3)}$, $g(x_3,\gamma_1,\gamma_2)$ is a linear function of
$x_3$, with positive and negative derivatives, respectively. Hence,
for given $\gamma_1,\gamma_2$, the minimum of
$g(x_3,\gamma_1,\gamma_2)$ is obtained in the interval
$x_3^{(3)}\leq x_3\leq x_3^{(4)}$. When $\gamma_2=0$,
$g(x_3,\gamma_1,\gamma_2)$ is an even function of $x_3$; its minimum
for given $\gamma_1$  is obtained at $x_3=0$ and is equal to
$\frac{1}{2}$. Otherwise, the partial derivative of
$g(x_3,\gamma_1,\gamma_2)$ with respect to $x_3$ is positive at
$x_3=0$; the minimum of $g(x_3,\gamma_1,\gamma_2)$ for given
$\gamma_1, \gamma_2$ is obtained in the interval $x_3^{(3)}\leq
x_3<0$ and is smaller than $\frac{1}{2}$.   Setting the derivative
of $g(x_3,\gamma_1,\gamma_2)$ with respect to $x_3$ to zero leads to
a fourth-order polynomial equation about $x_3$; the minimum can be
found after solving this equation. In particular, $-1<x_3<0$ at the
global minimum of $g(x_3,\gamma_1,\gamma_2)$. Figure~\ref{fig:gm3}
shows the dependence of the minimum of $g(x_3,\gamma_1,\gamma_2)$
over $x_3$ on $\gamma_1,\gamma_2$; the minimum is also the minimum
of $g(\rho_{\mathrm{rk}2})$ in the rank-two subspace. According to
the figure, the global minimum of $g(x_3,\gamma_1,\gamma_2)$ is
obtained in the rank-two subspace with
$\gamma_1=\gamma_2=\frac{\pi}{4}$.

To determine the maximally entangled  multipartite states is a
highly nontrivial task, since it usually involves a massive
optimization process over a large parameter space. Even for three
qubits, the maximally entangled state with respect to GM is not
known for sure, although it has been conjectured with strong
evidence that the W state is such a candidate \cite{twp09}. As an
immediate application of the above results, we prove this conjecture
rigorously in Appendix C.
\begin{theorem}
\label{W} Up to local unitary transformations, the W state is the
unique maximally entangled pure three-qubit state with respect to
GM.
\end{theorem}
Theorem~\ref{W} can be generalized to mixed states according to the
convex roof definition in Eq.~(\ref{eq:gmms}).
\begin{theorem}
\label{Wmixed} The W state is the maximally entangled state among
all three-qubit states with respect to GM.
\end{theorem}

\section{conclusions}

We have provided analytical methods for  deriving the GM of
symmetric pure multiqubit states with non-negative amplitudes in the
Dicke basis and that of symmetric pure three-qubit states. Also, we
have introduced a systematic method for  studying the GM of pure
three-qubit states in virtue of a canonical form of their bipartite
reduced states. In particular, we have derived explicit analytical
formulae of GM for the family of pure three-qubit states one of
whose rank-two two-qubit reduced states is a convex combination of
the maximally entangled state and its orthogonal pure state within
the rank-two subspace. Based on these results, we  further proved
that the W state is the maximally entangled three-qubit state with
respect to GM. Our studies can simplify the calculation of GM and
provide a better understanding of multipartite entanglement,
especially the entanglement in three-qubit states. Our results also
facilitate the comparison of GM with other entanglement measures,
like relative entropy of entanglement \cite{hmm08,zch10}. Moreover,
they may help investigate the physical phenomena in multipartite
entangled systems emerging in condensed matter physics.

\section*{Acknowledgement}
We thank Masahito Hayashi for critical reading of the manuscript. We
also thank Otfried G\"{u}hne for stimulating discussion on the paper
\cite{hkw09}. The Centre for Quantum Technologies is funded by the
Singapore Ministry of Education and the National Research Foundation
as part of the Research Centres of Excellence programme.  A. Xu is
supported by the program of Ningbo Natural Science Foundation
(2010A610099).

\section*{Appendix A: Derivation of Eqs.~(\ref{al:sytqaxiom1}--\ref{al:sytqaxiom3}) }

We will use the technique in Refs.~\cite{tpt08,tkk09} to simplify
the problem. According to the definition in Eq.~(\ref{eq:gm}),
\begin{align}
  \label{al:gsquare}
  G^2(\ket{\Ph}) &= \underset{ \ket{a} \ket{b} \ket{c}} {\mathrm{max} }
  \mathrm{Tr}\bigl[ \proj{\Ph} \bigl(\proj{a} \ox \proj{b} \ox \proj{c}\bigr )\bigr] \nonumber\\
           &= \underset{ \ket{a} \ket{b} } {\mathrm{max}}\;
  \mathrm{Tr}\bigl[ \bigl(\text{Tr}_C \proj{\Ph}\bigr) \bigl(\proj{a} \ox \proj{b}\bigr) \bigr],
\end{align}
where $\ket{a}, \ket{b}$ are normalized qubit states. The second
equality follows from Theorem 1 of E. Jung {\it et al.}
\cite{jung08}, which states that any ($n-1$)-qudit reduced state
uniquely determines the GM of the original $n$-qudit pure state, as
we have mentioned in the second paragraph of Sec.~\ref{sec:AP2}.

To reduce Eq.~(\ref{al:gsquare}), we use the Bloch sphere
representation of qubit \cite{nc00}:
\begin{equation}
  \label{eq:bloch}
  \r :=\frac12 \left( I + \bm{s}_{\r} \cdot \bm{\s} \right),
\end{equation}
where the components of $\bm{\s}$ are three Pauli matrices and
$\bm{s}_{\r}$ is the Bloch vector.

Suppose the states $\ket{a}, \ket{b}$ have  Bloch vectors
$\bm{s}_1,\bm{s}_2$ respectively. Then Eq.~(\ref{al:gsquare}) gives
rise to two sets of equations:
\begin{align}
  \label{al:twosets}
  &\bm{r}_1 + G \bm{s}_2 = \l_1 \bm{s}_1,\quad
  \bm{r}_2 + G \bm{s}_1 = \l_2 \bm{s}_2,\nonumber\\
  &\bm{r}_1 = \text{Tr}[\text{Tr}_{BC} (\proj{\Ph})\bm{\s}],\quad
  \bm{r}_2 = \text{Tr}[\text{Tr}_{AC} (\proj{\Ph})\bm{\s}],
\end{align}
where $\l_1, \l_2$ are Lagrange multipliers, and the $3 \oy 3$
matrix $G$ has elements $G_{ij}=\text{Tr}[(\text{Tr}_C
\proj{\Ph})(\s_i \ox \s_j)]$. Since the reduced density operators
$\text{Tr}_{BC} \proj{\Ph}$ and $\text{Tr}_{AC} \proj{\Ph}$ are
identical,  one can show that $\bm{r}_1=\bm{r}_2=\bm{r}$ after some
algebra. It follows that $\bm{s}_1=\bm{s}_2=\bm{s}$, $\l_1=\l_2=\l$,
and Eq.~(\ref{al:twosets}) reduces to
\begin{align}
  \label{al:oneset}
  \bm{r} + G \bm{s} = \l \bm{s}.
\end{align}
 The solutions to Eq.~(\ref{al:oneset}) determine the GM of
the state $\ket{\Ph}$ in Eq.~(\ref{eq:sytqgeneral}).

Define $\bm{s}=(\sin\t \cos\ph, \sin\t \sin\ph, \cos\t)$ with
$\t\in[0,\pi]$ and $\ph\in[0,2\pi]$; then Eq.~(\ref{al:oneset})
reduces to Eqs.~(\ref{al:sytqaxiom1})--(\ref{al:sytqaxiom3}).

\section*{ Appendix B: Derivation of Eq.~(\ref{gm1}) }
In this Appendix, we derive  Eq.~(\ref{gm1}). Recall that the
relevant parameter range is $0\leq
\gamma_2\leq\gamma_1\leq\frac{\pi}{2}$, $\gamma_2+\gamma_1\leq
\frac{\pi}{2}$,  and $-1\leq x_3\leq1$, see
Sec.~\ref{sec:canonical}. To simplify the following discussion, we
also assume $0<\gamma_1<\frac{\pi}{2}$ and $|x_3|<1$; but it turns
out that  the  final result is  applicable without this restriction.

 When $x_1=x_2=0$, according to Eqs.~(\ref{trace1}) and (\ref{fv1}),
\begin{eqnarray}
\label{fv2}
 f(a,b,c)&=&1+c\cos\gamma_1\cos\gamma_2+cx_3\sin\gamma_1\sin\gamma_2+|\bm{w}|,\nonumber\\
\bm{w}&=&\left(
             \begin{array}{c}
               a(\cos\gamma_2\sin\gamma_1- x_3\cos\gamma_1\sin\gamma_2) \\
               b(-x_3\cos\gamma_2\sin\gamma_1+\cos\gamma_1\sin\gamma_2) \\
               c x_3+x_3\cos\gamma_1\cos\gamma_2+\sin\gamma_1\sin\gamma_2\ \\
             \end{array}
           \right)^T.\nonumber\\
           \end{eqnarray}
According to Sec.~\ref{sec:generic}, to compute
$g(x_3,\gamma_1,\gamma_2)$, or equivalently, the maximum of
$f(a,b,c)$ over the unit sphere, we need only to maximize
$f_2(c)=f(\sqrt{1-c^2},0,c)$ over the single variable $c$ for
$-1\leq c\leq 1$. Here $f_2(c)$  can be expressed as follows,
\begin{eqnarray}\label{eq:f_2}
f_2(c)&=&1+u_0c+\sqrt{u_1c^2+2u_2c+u_3},\nonumber\\
u_0&=&\cos\gamma_1\cos\gamma_2+x_3\sin\gamma_1\sin\gamma_2>0,\nonumber\\
u_1&=&x_3^2-(\cos\gamma_2\sin\gamma_1-
x_3\cos\gamma_1\sin\gamma_2)^2,\nonumber\\
u_2&=&(x_3\cos\gamma_1\cos\gamma_2+\sin\gamma_1\sin\gamma_2)x_3,\nonumber\\
u_3&=&(\sin\gamma_1)^2+x_3^2(\cos\gamma_1)^2>0.
\end{eqnarray}
The four coefficients $u_0, u_1, u_2, u_3$ in Eq.~(\ref{eq:f_2})
satisfy the following relations,
\begin{eqnarray}\label{eq:u1234}
u_2> u_1,\quad u_2^2-u_1u_3>0,\quad u_0^2-u_1>0;
\end{eqnarray}
these relations are useful in the following discussion.

To determine the maximum of $f_2(c)$ for $-1\leq c\leq 1$, we shall
differentiate three cases according to the sign of $u_1$. Note that
$u_1$ is a quadratic function of $x_3$ with a positive quadratic
coefficient, and that it has the following two zeros:
\begin{eqnarray}
&&x_3^{(1,2)}=\frac{\cos\gamma_2\sin\gamma_1}{\pm1+\cos\gamma_1\sin\gamma_2},\label{interval1}
\end{eqnarray}
which satisfy the  inequalities: $-1\leq x_3^{(2)}<0< x_3^{(1)}<1$;
 $x_3^{(2)}$ is equal to $-1$ only if $\gamma_1+\gamma_2=1$.

\textbf{Case 1:} $x_3=x_3^{(1)}$ or $x_3=x_3^{(2)}$. In this case
$u_1=0$, $u_0, u_2>0$,
\begin{eqnarray}
f_2(c)&=&1+u_0c+\sqrt{2u_2c+u_3},
\end{eqnarray}
so the maximum of $f_2(c)$ can only be obtained at $c=1$.

\textbf{Case 2:} $x_3<x_3^{(2)}$ or $x_3>x_3^{(1)}$. In this case,
$u_0, u_1, u_2>0$, the discriminant of the quadratic function
$u_1c^2+2u_2c+u_3$ about $c$ is $4(u_2^2-u_1u_3)>0$, so the
quadratic function  has two zeros with mean $-u_2/u_1<0$. Since the
quadratic function must be nonnegative in the interval $[-1,1]$ by
definition, both  zeros must be smaller than or equal to $-1$. In
the interval $[-1,1]$, this quadratic function and $f_2(c)$ are both
strictly  increasing, so the maximum of $f_2(c)$ can only be
obtained at $c=1$.

 \textbf{Case 3:} $x_3^{(2)}< x_3<x_3^{(1)}$. In this
case $u_1<0$, the quadratic function $u_1c^2+2u_2c+u_3$ is positive
between its two zeros. One zero is smaller than or equal to $-1$,
and the other larger than or equal to 1. To determine the maximum of
$f_2(c)$, we take the first and the second derivatives of $f_2(c)$:
\begin{eqnarray}
f_2^\prime(c)&=&u_0+\frac{u_1c+u_2}{\sqrt{u_1c^2+2u_2c+u_3}},\nonumber\\
f_2^{\prime\prime}(c)&=&\frac{u_1u_3-u_2^2}{(u_1c^2+2u_2c+u_3)^{3/2}}<0,
\end{eqnarray}
where the last inequality follows from Eq.~(\ref{eq:u1234}). There
is only one solution to the equation $f_2^\prime(c)=0$,
\begin{eqnarray}
\bar{c}&=&-\bigl[x_3^2-(\cos\gamma_2\sin\gamma_1-x_3\cos\gamma_1\sin\gamma_2)^2\bigr]^{-1}\nonumber\\
&&\times\bigl[x_3(x_3\cos\gamma_1\cos\gamma_2+\sin\gamma_1\sin\gamma_2)\nonumber\\
&&+\sin\gamma_1(\cos\gamma_1\cos\gamma_2+x_3\sin\gamma_1\sin\gamma_2)\nonumber\\
&&\times(\sin\gamma_1-x_3\cos\gamma_1\tan\gamma_2)\bigr]\geq0.\label{c1}
\end{eqnarray}
Since the second derivative of $f_2(c)$ is always negative,
$\bar{c}$ is the global maximum of the function $f_2(c)$ in the
interval where it is real valued. Restricted to the interval
$[-1,1]$, the maximum of $f_2(c)$ is obtained at $\bar{c}$ if
$\bar{c}<1$ and at $c=1$ otherwise. In both cases, the maximum point
is unique. Hence, it remains to determine when $\bar{c}\geq1$ and
when $\bar{c}< 1$.

After some algebra, one can show that $x_3^{(1,2)}$ defined in
Eq.~(\ref{interval1}) and $x_3^{(3,4)}$ defined in
Eq.~(\ref{interval2}) satisfy the following inequalities,
\begin{eqnarray}
-1\leq x_3^{(2)}\leq x_3^{(3)}<0\leq x_3^{(4)}< x_3^{(1)}<1.
\end{eqnarray}
If $\gamma_1+\gamma_2=\frac{\pi}{2}$, then $x_3^{(3)}=x_3^{(2)}=-1$,
and $\bar{c}$ satisfies the following relation,
\begin{eqnarray}
\left\{ \begin{array}{cl}
         \bar{c}<1   &\quad -1<x_3< x_3^{(4)}, \\
         \bar{c}\geq 1      &\quad x_3^{(4)}\leq x_3<x_3^{(1)}.
       \end{array}\right.
\end{eqnarray}
If $\gamma_1+\gamma_2<\frac{\pi}{2}$, then $-1<x_3^{(2)}<x_3^{(3)}$,
and $\bar{c}$ satisfies the following relation,
\begin{eqnarray}
\left\{ \begin{array}{cl}
         \bar{c}<1   &\quad x_3^{(3)}<x_3<x_3^{(4)}, \\
         \bar{c}\geq 1      &\quad x_3^{(2)}<x_3\leq x_3^{(3)}\quad \mbox{or} \quad
         x_3^{(4)}\leq x_3<x_3^{(1)}.
       \end{array}\right.
\end{eqnarray}

According to the observations in the above three cases, if
$-1<x_3\leq x_3^{(3)}$ or $x_3^{(4)}\leq x_3<1$, the maximum of
$f_2(c)$ is obtained at 1; if $-x_3^{(3)}<x_3< x_3^{(4)}$, the
maximum is obtained at $\bar{c}$. The maximum point is unique in
both cases. The values of $f_2(c)$ at 1 and $\bar{c}$ are
respectively given by
\begin{eqnarray}
&&f_2\bigl(\bar{c}\bigr)=\nonumber\\
&&\frac{2(1-x_3^2)\sin\gamma_1\cos\gamma_2(\cos\gamma_2\sin\gamma_1-x_3\cos\gamma_1\sin\gamma_2)}
{-\bigl[x_3^2-(\cos\gamma_2\sin\gamma_1-x_3\cos\gamma_1\sin\gamma_2)^2\bigr]},\nonumber\\
&&f_2(1)=1+\cos\gamma_1\cos\gamma_2+x_3\sin\gamma_1\sin\gamma_2\nonumber\\
&&\hphantom{f_2(1)=}+|x_3+x_3\cos\gamma_1\cos\gamma_2+\sin\gamma_1\sin\gamma_2|\nonumber\\
&&\hphantom{f_2(1)}=\left\{
\begin{array}{cc}
(1-x_3)[1+\cos(\gamma_1+\gamma_2)],&\quad x_3\leq
x_3^{(3)},\\
(1+x_3)[1+\cos(\gamma_1-\gamma_2)], &\quad  x_3\geq x_3^{(4)},
\end{array}
\right.
\end{eqnarray}
where in deriving the last equality, we have noticed that
\begin{eqnarray}
\left\{\begin{array}{cc}
x_3+x_3\cos\gamma_1\cos\gamma_2+\sin\gamma_1\sin\gamma_2 \leq 0, & \quad x_3\leq x_3^{(3)}, \\
x_3+x_3\cos\gamma_1\cos\gamma_2+\sin\gamma_1\sin\gamma_2  \geq 0,
&\quad  x_3\geq x_3^{(4)}.
\end{array}\right.\nonumber
\end{eqnarray}
Now Eq.~(\ref{gm1}) follows immediately when
$0<\gamma_1<\frac{\pi}{2}$ and $|x_3|<1$; recall that
$g(x_3,\gamma_1,\gamma_2)=\frac{1}{4}\max_{-1\leq c\leq 1} f_2(c)$.
It is straightforward to verify that the formula is also valid in
the special cases $|x_3|=1$ or $\gamma_1=0,\frac{\pi}{2}$, hence the
derivation is complete.

\section*{Appendix C: Proof of Theorem~\ref{W}}
To prove Theorem~\ref{W} in Sec.~\ref{sec:W}, it suffices to show
that the global minimum of $g(\rho_{\mathrm{rk}2})$ is obtained at
the two-qubit reduced state of the W state.

In the relevant parameter range   $0\leq \gamma_2\leq
\gamma_1\leq\frac{\pi}{2}$, $\gamma_1+\gamma_2\leq \frac{\pi}{2}$,
by taking its derivative with respect to $\gamma_2$ in
Eq.~(\ref{gm1}), one can show that, for given $\gamma_1, x_3$,
$g(x_3,\gamma_1,\gamma_2)$ is monotonically decreasing with
$\gamma_2$ for $x_3<0$ and monotonically increasing with $\gamma_2$
for  $x_3>0$. Assuming $x_3<0$, where the global minimum point
should satisfy according to Sec.~\ref{sec:W}; then
$g(x_3,\gamma_1,\gamma_2)$ is monotonically decreasing with
$\gamma_2$. Hence, either $\gamma_1=\gamma_2$ or
$\gamma_1+\gamma_2=\frac{\pi}{2}$ at the global minimum of
$g(x_3,\gamma_1,\gamma_2)$. We shall show that the unique minimum is
obtained at the two-qubit  reduced state of the W state in both
cases.

\subsubsection{special case: $\gamma_1+\gamma_2=\frac{\pi}{2}$}
If $\gamma_2=\frac{\pi}{2}-\gamma_1$ ($\frac{\pi}{4}\leq
\gamma_1\leq \frac{\pi}{2}$),
 $\rho_{\mathrm{rk}2}$ is supported on the
symmetrical subspace, according to
Eqs.~(\ref{eq:r2projector})--(\ref{eq:Pauli}). In this case,
Eq.~(\ref{gm1}) reduces to
\begin{eqnarray}
 &&g(x_3,\gamma_1,\frac{\pi}{2}-\gamma_1)=\nonumber\\
 &&\left\{
 \begin{array}{cl}
   \frac{1}{2}-\frac{(1+x_3)x_3(\cos\gamma_1)^2}{-1+3x_3+(1+x_3)\cos(2\gamma_1)}, &
   -1\leq x_3 < x_3^{(4)},\vspace{0.5ex}\\
      \frac{(1+x_3)[1+\sin(2\gamma_1)]}{4}, &x_3^{(4)}\leq x_3\leq1,
 \end{array}\right.
\end{eqnarray}
where
\begin{eqnarray}
&&x_3^{(4)}=\frac{1-\sqrt{2}\sin(2\gamma_1+\frac{\pi}{4})}{3+\sqrt{2}\sin(2\gamma_1+\frac{\pi}{4})}.
\end{eqnarray}
$g(x_3,\gamma_1,\frac{\pi}{2}-\gamma_1)$ is equal to $\frac{1}{2}$
at $x_3=0,-1$, independent of $\gamma_1$;
$g(x_3,\gamma_1,\frac{\pi}{2}-\gamma_1)$ is monotonically increasing
with $\gamma_1$ for $-1<x_3<0$, and monotonically decreasing for
$0<x_3\leq1$ (see also Fig.~\ref{fig:gm0}).

To determine the maximum of $g(x_3,\gamma_1,\frac{\pi}{2}-\gamma_1)$
for given $\gamma_1$, we can set  its derivative with respect to
$x_3$ to 0 (in the interval $-1\leq x_3 < x_3^{(4)}$), which leads
to the following quadratic equation over $x_3$,
\begin{eqnarray}
&&[3+\cos(2\gamma_1)]x_3^2+[-2+2\cos(2\gamma_1)]x_3+\cos(2\gamma_1)=1.\nonumber\\
\end{eqnarray}
Given $\frac{\pi}{4}\leq \gamma_1\leq\frac{\pi}{2}$, the only
solution with modulus less than or equal to 1 is
\begin{eqnarray}
x_3^{(5)}=\frac{2\sin\gamma_1(\sin\gamma_1-\sqrt{2})}{3+\cos(2\gamma_1)}.
\end{eqnarray}
The minimum of $g(x_3,\gamma_1,\frac{\pi}{2}-\gamma_1)$ for given
$\gamma_1$ is
\begin{eqnarray}
g(x_3^{(5)},\gamma_1,\frac{\pi}{2}-\gamma_1)=\frac{[1+\cos(2\gamma_1)+\sqrt{2}\sin\gamma_1]^2}{[3+\cos(2\gamma_1)]^2}.
\end{eqnarray}
One can show  that $g(x_3^{(5)},\gamma_1,\frac{\pi}{2}-\gamma_1)$ is
monotonically increasing with respect to $\gamma_1$ by taking its
derivative with respect to $\gamma_1$;  hence, its minimum is
obtained at $\gamma_1=\frac{\pi}{4}$. At this minimum point,
$\gamma_1=\gamma_2=\frac{\pi}{4}$, $x_3=x_3^{(5)}=-\frac{1}{3}$, and
$g(-\frac{1}{3},\frac{\pi}{4},\frac{\pi}{4})=\frac{4}{9}$. This
minimum is also the global minimum of $g(\rho_{\mathrm{rk}2})$.

The rank-two state corresponding to this minimum is exactly the
two-qubit reduced state of the W state, moreover, up to local
unitary transformations, the W state is the only pure three-qubit
state with this rank-two state as a two-qubit reduced state. To see
this, recall that the two-qubit reduced state of
$|\mathrm{W}\rangle=\frac{1}{\sqrt{3}}(|100\rangle+|010\rangle+|001\rangle)$
is
\begin{eqnarray}
\rho_{\mathrm{rk}2}(\mathrm{W})&=&\frac{1}{3}|00\rangle\langle00|+\frac{2}{3}|\psi^+\rangle\langle\psi^+|
\end{eqnarray}
with $|\psi^+\rangle=\frac{1}{\sqrt{2}}(|01\rangle+|10\rangle)$,
which is  a convex combination of a pure product state and a
orthogonal  Bell state, thus $\gamma_1=\gamma_2=\frac{\pi}{4}$,
according to Ref.~\cite{em02}. In this rank-two subspace, the Bloch
vectors of the two states $|00\rangle$ and $|\psi^+\rangle$ are
$(0,0,1)$ and $(0,0,-1)$ respectively, and the Bloch vector of the
state $\rho_{\mathrm{rk}2}(\mathrm{W})$ is exactly
$(0,0,-\frac{1}{3})$.

\subsubsection{special case: $\gamma_2=\gamma_1\leq \frac{\pi}{4}$}
In this case, Eq.~(\ref{gm1}) reduces to
\begin{eqnarray}
&&g(x_3,\gamma_1,\gamma_1)=\nonumber\\
&&\left\{
\begin{array}{cl}
  \frac{(1-x_3)}{2}(\cos\gamma_1)^2,& -1\leq x_3\leq x_3^{(3)},\vspace{0.5ex}\\
 \frac{-(1-x_3)^2(1+x_3)[\sin(2\gamma_1)]^2}{-1+x_3(2+7x_3)+(1-x_3)^2\cos(4\gamma_1)},& x_3^{(3)}< x_3<0, \vspace{0.5ex}\\
  \frac{(1+x_3)}{2} ,& 0\leq x_3\leq1,
\end{array}
\right.\nonumber\\
\end{eqnarray}
where
\begin{eqnarray}
&&x_3^{(3)}=-(\tan\gamma_1)^2.
\end{eqnarray}
It is interesting to note that $g(x_3,\gamma_1,\gamma_1)$ is
independent of $\gamma_1$ when $0\leq x_3\leq1$. If $x_3<0$,
$g(x_3,\gamma_1,\gamma_1)$ is monotonically decreasing with
$\gamma_1$. Hence,  $\gamma_1=\frac{\pi}{4}$ at its minimum,  the
corresponding rank-two subspace is then symmetric. According to the
result on symmetric states in the previous subsection, the unique
minimum of $g(x_3,\gamma_1,\gamma_1)$ is also obtained at
$\gamma_1=\gamma_2=\frac{\pi}{4}$, $x_3=-\frac{1}{3}$.

We have shown that the unique minimum of $g(x_3,\gamma_1,\gamma_2)$
is obtained at $\gamma_1=\gamma_2=\frac{\pi}{4}$,
$x_3=-\frac{1}{3}$, and that the corresponding state is the
two-qubit reduced state of the W state. This minimum is also the
global minimum of $g(\rho_{\mathrm{rk}2})$. To prove that the
minimum is unique among all two-qubit rank-two states, it remains to
show that it is unique in the rank-two subspace with
$\gamma_1=\gamma_2=\frac{\pi}{4}$. It suffices to verify that
$g(x_1,x_2,-\frac{1}{3},\frac{\pi}{4},\frac{\pi}{4})>g(0,0,-\frac{1}{3},\frac{\pi}{4},\frac{\pi}{4})$
for $x_1^2+x_2^2>0$ (here we write
$g(x_1,x_2,x_3,\gamma_1,\gamma_2)$ for $g(\rho_{\mathrm{rk}2})$).
Due to the rotational symmetry of $g(x_1,x_2,x_3,\gamma_1,\gamma_2)$
about the $x_3$ axis discussed in Sec.~\ref{sec:canonical}, this is
true if
$g(x_1,0,-\frac{1}{3},\frac{\pi}{4},\frac{\pi}{4})>g(0,0,-\frac{1}{3},\frac{\pi}{4},\frac{\pi}{4})$
for $x_1>0$. According to Eq.~(\ref{fv1}), when
$\gamma_1=\gamma_2=\frac{\pi}{4}$, $x_2=0$, $x_3=-\frac{1}{3}$,
\begin{eqnarray}
&&f\Bigl(\frac{2\sqrt{2}}{3},0,\frac{1}{3}\Bigr)=\frac{2}{9}\bigl[5+3x_1+\sqrt{9+3x_1(10+9x_1)}\,\bigr];\nonumber\\
\end{eqnarray}
hence,
\begin{eqnarray}
&&g\Bigl(x_1,0,-\frac{1}{3},\frac{\pi}{4},\frac{\pi}{4}\Bigr)\geq
\frac{1}{4}f\Bigl(\frac{2\sqrt{2}}{3},0,\frac{1}{3}\Bigr),\nonumber\\
&&
{}>\frac{4}{9}=g\Bigl(0,0,-\frac{1}{3},\frac{\pi}{4},\frac{\pi}{4}\Bigr).
\end{eqnarray}
This  completes the proof of Theorem~\ref{W}.


\begin{thebibliography}{99}


\bibitem{epr35} A. Einstein, B. Podolsky, N. Rosen, Phys. Rev. {\bf47}, 777 (1935).

\bibitem{schrodinger35} E. Schr\"odinger, Naturwissenschaften, {\bf23}, 807 (1935).

\bibitem{hhh09} R. Horodecki, P. Horodecki, M. Horodecki, K. Horodecki, Rev. Mod. Phys. {\bf81}, 865 (2009).

\bibitem{rb01} R. Raussendorf and H. J. Briegel, Phys. Rev. Lett. \textbf{86}, 5188
(2001); For a review on one-way quantum computation, see E. Campbell
and J. Fitzsimons, arXiv:0906.2725 [quant-ph].

\bibitem{lgg09} C. Y.  Lu, W. B. Gao, O. G\"uhne, X. Q. Zhou, Z. B. Chen,
and J. W. Pan, Phys. Rev. Lett. {\bf102}, 030502 (2009).

\bibitem{hbb99} M. Hillery, V. Bu\v{z}ek, and A. Berthiaume, Phys. Rev. A {\bf 59}, 1829 (1999).


\bibitem{spc09} V. Scarani, H. B. Pasquinucci, N. J. Cerf,
M. Du\^{a}sek, N. L\"{u}tkenhaus, and M. Peev, Rev. Mod. Phys.
{\bf81}, 001301 (2009).

\bibitem{bbc93}  C. H. Bennett, G. Brassard, C. Cr\'{e}peau, R.
Jozsa, A. Peres, and W. K. Wootters, Phys. Rev. Lett. {\bf 70}, 1895
(1993).

\bibitem{ghz89} D. M. Greenberger, M. A. Horne, and A. Zeilinger, in {\it Bell's
Theorem, Quantum Theory, and Conceptions of the Universe\/}, edited
by M. Kafatos (Kluwer, Dordrecht, 1989), p. 69.

\bibitem{ghsz90}D. M. Greenberger, M. A. Horne, A. Shimony, and A.
Zeilinger, Am. J. Phys. {\bf 58}, 1131 (1990).

\bibitem{bbg09} J. D. Bancal, C. Branciard, N. Gisin, and
S. Pironio, Phys. Rev. Lett. {\bf103}, 090503 (2009).


\bibitem{ah09} R. Augusiak and P. Horodecki, Phys. Rev. A {\bf80}, 042307 (2009).

\bibitem{lzg07}C. Y. Lu,
X. Q. Zhou, O. G\"uhne, W. B. Gao, J. Zhang, Z. S. Yuan, A. Goebel,
 T. Yang, and J. W. Pan, Nat. Phys. {\bf3}, 91 (2007).

\bibitem{pcd09} S. B. Papp, K. S. Choi, H. Deng, P. Lougovski,
S. J. van Enk, H. J. Kimble, Science {\bf324}, 764 (2009).

\bibitem{dicke54} R. H. Dicke, Phys. Rev. {\bf93}, 99 (1954).

\bibitem{wkk09} W. Wieczorek, R. Krischek, N. Kiesel,
P. Michelberger, G. Toth, and H. Weinfurter, Phys. Rev. Lett.
{\bf103}, 020504 (2009).

\bibitem{pct09} R. Prevedel, G. Cronenberg, M. S. Tame,
M. Paternostro, P. Walther, M. S. Kim, and A. Zeilinger, Phys. Rev.
Lett. {\bf103}, 020503 (2009).

\bibitem{kkb09} P. Krammer, H. Kampermann, D. Bruss, R. A. Bertlmann,
L. C. Kwek, and C. Macchiavello, Phys. Rev. Lett. {\bf103}, 100502
(2009).

\bibitem{vidal00} G. Vidal, J. Mod. Opt. {\bf47}, 355
(2000).

\bibitem{bds96} C. H. Bennett, D. P. DiVincenzo, J. A. Smolin, and W. K. Wootters,
Phys. Rev. A {\bf54}, 3824 (1996).

\bibitem{pv07} For a review, see M. B. Plenio and S. Virmani, Quantum Inf. Comput. {\bf7}, 1 (2007).

\bibitem{nc00}  M. A. Nielsen and I. L. Chuang, {\it Quantum Computation
and Quantum Information} (Cambridge University Press, Cambridge,
England, 2000).

\bibitem{wg03} T.-C. Wei and P. M. Goldbart, Phys. Rev. A {\bf68}, 042307 (2003).

\bibitem{bh01} D. C. Brody, and L. P. Hughston, J. Geom. Phys.
{\bf38}, 19 (2001). The paper proposed the geometric measure of
entanglement for bipartite pure states in an earilier version than
\cite{wg03}.

\bibitem{mw02} D. A. Meyer and N. R. Wallach, J. Math. Phys. {\bf43}, 4273 (2002).

\bibitem{ckw00}V. Coffman, J. Kundu and W. K. Wootters, Phys. Rev.
A {\bf61}, 052306 (2000).

\bibitem{twp09} S. Tamaryan, T.-C. Wei and D. Park, Phys. Rev. A {\bf 80}, 052315 (2009).


\bibitem{ffp08} P. Facchi, G. Florio, G. Parisi, and
S. Pascazio, Phys. Rev. A {\bf77}, 060304R (2008).

\bibitem{hmm08} M. Hayashi, D. Markham, M. Murao, M. Owari, and S. Virmani, Phys.
Rev. A {\bf77}, 012104 (2008).

\bibitem{hmm06} M. Hayashi, D. Markham, M. Murao, M. Owari, and S. Virmani, Phys.
Rev. Lett. {\bf96}, 040501 (2006).

\bibitem{gfe09} D. Gross, S. T. Flammia, and J. Eisert,
Phys. Rev. Lett. {\bf102}, 190501 (2009).

\bibitem{oru08}R. Or\'us, Phys. Rev. Lett. {\bf 100}, 130502 (2008).

\bibitem{odv08} R. Or\'us, S. Dusuel, and J. Vidal, Phys. Rev. Lett. {\bf101}, 025701 (2008).

\bibitem{oru08b}R. Or\'us, Phys. Rev. A {\bf 78}, 062332 (2008).

\bibitem{ow09}R. Or\'us, and T.-C. Wei, arXiv:0910.2488 [cond-mat.str-el].

\bibitem{tkk09} S. Tamaryan, H. Kim, M. S. Kim, K. S. Jang and D. K. Park, arXiv:0909.1077 [quant-ph].

\bibitem{tpt08} L. Tamaryan, D. K. Park and S. Tamaryan, Phys. Rev. A {\bf 77}, 022325
(2008).


\bibitem{hs09} J. J. Hilling and A. Sudbery, J. Math. Phys. {\bf 51}, 072102 (2010).

\bibitem{pr09} P. Parashar, S. Rana, arXiv:0909.4443 [quant-ph].

\bibitem{chen09} X. Y. Chen, J. Phys. B: At. Mol. Opt. Phys. {\bf 43}, 085507 (2010).


\bibitem{mmm09} M. Hayashi, D. Markham, M. Murao, M. Owari, and S.
Virmani, J. Math. Phys. {\bf 50}, 122104 (2009).

\bibitem{hkw09} R. H\"{u}bener, M. Kleinmann, T. -C. Wei, C. G. Guill\'{e}n, and O. G\"{u}hne,
Phys. Rev. A {\bf80}, 032324 (2009).

\bibitem{sb07} Y. Shimoni and O. Biham, Phys. Rev. A {\bf75}, 022308 (2007).

\bibitem{msb10} Y. Most, Y. Shimoni, O. Biham, Phys. Rev. A {\bf81}, 052306 (2010).

\bibitem{em02} B.-G. Englert, N. Metwally, Kinematics of qubit pairs, in {\it Mathematics of
quantum computation}, edited by R. K. Brylinski, G. Chen (Chapman \&
Hall/CRC Press, Boca Raton, 2002).


\bibitem{zch10} Huangjun Zhu, Lin Chen, and Masahito Hayashi, New J. Phys. {\bf 12}, 083002
(2010).

\bibitem{notation1} Actually only two cases $\g=0, \frac{\pi}{2}$ were
handled in Ref. \cite{tkk09}. To compute the GM for the state
$\ket{\Ph}$ with $\g=-\frac{\pi}{2}$, it suffices to perform the
phase gate $\s_z \ox \s_z \ox \s_z$ on the state $\ket{\Ph}$. So the
states $\ket{\Ph}$ with $\g=-\frac{\pi}{2}$ and $\g=\frac{\pi}{2}$,
respectively, have the same GM.

\bibitem{jung08} E. Jung, M.-R. Hwang, H. Kim, M.-S. Kim, D. K. Park, J.-W. Son, and S. Tamaryan,
Phys. Rev. A {\bf77}, 062317 (2008).







\end{thebibliography}
\end{document}